\documentclass[twocolumn,showpacs,showkeys,preprintnumbers,superscriptaddress,floatfix,nofootinbib]{revtex4}
\usepackage{latexsym,amscd,amssymb,amsopn,amsthm,amsmath}
\usepackage{graphicx,color}
\newcounter{nnn}

\begin{document}
\title{SOME PRELIMINARY RESULTS ON RELATION BETWEEN TRIPLET COMPOSITION AND TISSUE SOURCE IN LARCH TOTAL TRANSCRIPTOME}
\author{Michael G.\,Sadovsky}
\affiliation{Institute of computational modelling of SD of RAS;\\ 660036 Russia, Krasnoyarsk,
Akademgorodok.} \email{msad@icm.krasn.ru}
\affiliation{Siberian Federal university;\\ 660041 Russia, Krasnoyarsk,
Svobodny prosp.79.}
\author{Tatiana A.\,Guseva}
\affiliation{Siberian Federal university;\\ 660041 Russia, Krasnoyarsk,
Svobodny prosp.79.}\email{dianema2010@mail.ru}
\author{Vladislav V.\,Birukov}
\affiliation{Siberian Federal university;\\ 660041 Russia, Krasnoyarsk,
Svobodny prosp.79.} \email{vladbir2010@gmail.com}
\author{Tatiana O.\,Shpagina}
\affiliation{Siberian Federal university;\\ 660041 Russia, Krasnoyarsk,
Svobodny prosp.79.}\email{shpagusa@mail.ru}
\author{Victoria D.\,Fedotovskaya}
\affiliation{Siberian Federal university;\\ 660041 Russia, Krasnoyarsk,
Svobodny prosp.79.}\email{pobedaXXXX@gmail.com}

\begin{abstract}
We studied the structuredness ensemble of transcriptome of Siberian larch. The clusters in 64-dimensional space were identified with $K$-means technique, where the objects to be clusterized are the different fragments of the genome. A tetrahedron like structure in distribution of these fragments was found. Chargaff's discrepancy measure was determined for each class, as well as that latter between the classes. It reveals a relative similitude of the classes. The results have been compared to those obtained for specific transcriptome of each tissue. Also, a surrogate transcriptome has been developed comprising the contigs assembled for specific tissues; that latter has been compared with the real total transcriptome, and significant difference has been observed.
\end{abstract}
\keywords{frequency; triplet; order; cluster; elastic map; evolution}

\pacs{87.10.+e, 87.14.Gg, 87.15.Cc, 02.50.-r}

\maketitle

\renewcommand{\geq}{\geqslant}
\renewcommand{\leq}{\leqslant}
\renewcommand{\baselinestretch}{1}
\section{Introduction}
Fast progress in technologies of DNA sequencing results in tremendous growth of the data available for analysis. Moreover, sequencing techniques go ahead of other ones devoted to annotation and deeper analysis of the deciphered sequences. Such bulk quantities of data challenge researchers in invention and implementation of some novel and non-conventional approaches and techniques to retrieve knowledge and order in those bulky data. Here we explore such approach in analysis of some genetic data collected Labouratory of forest genomics of Siberian federal university, under the project on Siberian larch genome deciphering, namely, the ensemble of contigs of transcriptome of \textit{L.\,sibirica}~Ledeb.

Total transcriptome is an important object in various researches in bioinformatics. That latter is the transcriptome (i.\,e. an entire ensemble of all the genes in an organism transcribed at the given time moment) with neither respect to the specific source of the genetic matter to be expressed. Thus, a total transcriptome (as a product of sequencing and further assembling) is a matter of interference of the reads that have been copied from different issues of an organism to be deciphered. At the first glance, such genetic object seems to be rather artificial: indeed, why one should assemble a transcriptome from the reads provided from different tissues, if it is possible to to do it separately.

Actually, a possibility to isolate a specific tissue for sampling is matter of luck. Quite often, there is no way to get a properly isolated sample, and deterioration is inevitable. Thus, a question arises towards the impact of such deterioration on assembling and further annotation of a transcriptome. In such capacity, a comprehensive study of a total transcriptome of an organism elucidates the effects that may come from the deterioration mentioned above, and be a kind of a reference, in the estimation of the assembling results, for specific cases.

Let now introduce some basic notions and concepts. Let $\mathfrak{T}$ be a sequence from four-letter alphabet $\aleph = \{\mathsf{A},\mathsf{C},\mathsf{G},\mathsf{T}\}$; biologically, it corresponds to DNA sequence of some nature. It means that $\mathfrak{T}$ may correspond to a genome, to a gene, to a part of gene, to a contig, etc., depending on the specific goal of a research. The length $N=|\mathfrak{T}|$ of a sequence is just the number of nucleotides in it. The details on that subject see below.

Triplet frequency dictionary is the key entity, in our study. A triplet frequency dictionary $W_{(3,t)}$ is the list of all 64 triplets counted within a sequence, where the window of a triplet identification moves upright (for certainty) over a sequence with the step~$t$. Parameter~$t$ is a matter of choice, and depends on the specific task to be solved. Here we shall consider the frequency dictionary~$W_{(3,1)}$, only. In other words, the dictionary of the first type enlists all the triplets met in a sequence, and any nucleotide yields a start of a triplet. Strictly speaking, to hold this definition true, one must connect the sequence under consideration into a circle, meanwhile that is not important for further presentation; see details in \cite{slovar1,slovar2,slovar3,sadov1,sadov3}.



\section{Genetic material}
Sequencing of \textit{L.\,sibirica}~Ledeb. total transcriptome was carried out in Labouratory of forest genomics of Siberian federal university. There were obtained four groups of tissue specific read ensembles: from needles, from cambium, from shoot, and from seedling. These four read ensembles have been assembled separately; also, the total transcriptome has been assembled through the merge
\begin{figure}
\includegraphics[width=0.49\textwidth]{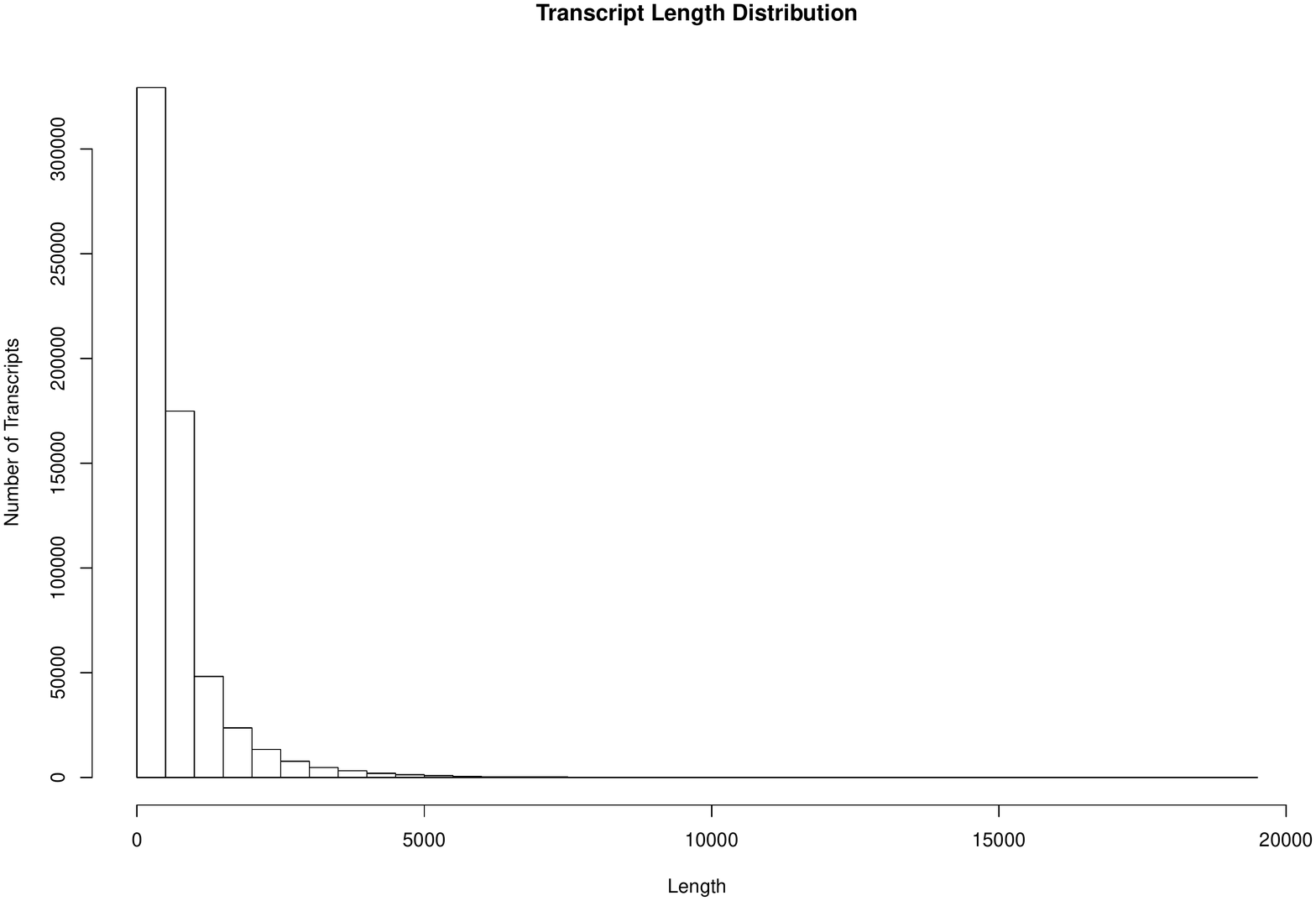}
\caption{\label{gist}The histogram of the number of contigs with the given length, for total transcriptome.}
\end{figure}

Here we shall consider the ensemble of contigs of \textit{L.\,sibirica}~Ledeb. transcriptome. Actually, we shall consider the following data sources:
\begin{list}{--}{\leftmargin=6mm \labelwidth=5mm \topsep=0mm \labelsep=2mm \itemsep=1pt \parsep=0mm \itemindent=10pt}
\item transcriptome of needles;
\item seedling transcriptome;
\item shoot transcriptome;
\item cambium transcriptome, and finally
\item the total transcriptome of \textit{L.\,sibirica}~Ledeb. and surrogate total transcriptome of \textit{L.\,sibirica}~Ledeb.
\end{list}
Real transcriptomes comprise the contigs of various lengths. Some figures characterizing the specific (as well, as the total one) transcriptomes are shown in Table~\ref{dl}; the table presents the figures for the shortest contig ($L_{\min}$), for the longest contig ($L_{\max}$), average length of transcriptome ($\langle L\rangle$), and total abundance of contigs in a transcriptome ($M$). The distribution of the contigs over the lengths is shown in Fig.~\ref{gist}.

For the proposes of the clustering and analysis of transcriptomes, we selected the subsets of contigs, in each specific transcriptome (including the total one). We took into the subsets sufficiently long contigs, only. The idea standing behind such selection is following: shorter contigs would yield rather abundant subsets of points (in 64-dimensional space) that are in local quasi-equilibrium: in other words, too many short contigs would have zero frequency of some triplets. Moreover, a greater number of triplets would be presented in a single copy, in a number of such shorter contigs, thus yielding a kind of quasi-equilibrium over the subspace determined by these triplets.

To avoid the above mentioned effect, we have eliminated shorter contigs. We comprise sufficiently long contigs, to carry out clustering and visualization of the data. Table~\ref{dl} shows the figures used to select the contigs involved into analysis: $L_d$ is the cut-off length of the contigs, in each specific transcriptome. That former means that we selected the contigs longer than~$L_d$; $M_d$ figures show the abundances of the sets of selected longer contigs.

To gain the total transcriptome, the reads ensembles obtained for each specific tissue have been merged into a single ensemble, and assembling has been carried out \cite{RAHMAN201817,10.1371/journal.pone.0050226}. Common idea in total transcriptome implementation is to enforce the coverage level of the genes expressed in various tissues, thus improving assembling of \textsl{de novo} sequence. Not discussing here an efficiency (quite arguable, frankly speaking), we just stress that a total transcriptome still is a good first step, in any genome deciphering being a kind of \textit{mean filed} approximation.

To evaluate a contribution of each specific tissue transcriptome into the total one, we implemented a surrogate transcriptome. That latter has been obtained by merging of the transcriptomes of each specific tissue into a general one. To do it, we used the longer contigs selected within each specific transcriptome.

All genetic material was obtained at Labouratory of forest genomics of Siberian federal university.

\begin{table}
\caption{\label{dl}Some figures characterizing transcriptomes. $L_{\min}$ is the minimal contig length, $L_{\max}$ the maximal contig length, $\langle L\rangle$ is average contig length, $L_d$ is the selection length, and $M_d$ is the abundance of contig set taken into consideration.}
\begin{ruledtabular}
\begin{tabular}{lcrcccrc}
transcriptome &$L_{\min}$&$L_{\max}$ &$\langle L\rangle$ &$L_d$&$M$&$M_d$\\\hline
needles & 201 & 9880 & 364 & 1000 & 59317 & 1851\\
shoot & 201 & 17893 & 532 & 5000 & 590240 & 1754\\
seedlings & 201 & 11008 & 455 & 2500 & 174805 & 1943\\
cambium & 201 & 20596 & 497 & 5000 & 628197 & 1455\\\hline
total & 301 & 19274 & 739 & 4000 & 610954 & 5713\\\\
\end{tabular}
\end{ruledtabular}
\end{table}

\section{Clustering methods and visualization}\label{strapglav}
We used freely distributed software \textsl{ViDaExpert} by Andrew Zinovyev (\url{bioinfo.curie.fr}) for visualization data. Also, $K$-means clustering technique \cite{dyn1} has been applied, to prove a structuredness in transcriptome data.

Let now explain the visualization technique. To retrieve a structure pattern in transcriptome (any of them, enlisted above), each contig was converted into frequency dictionary~$W_{(3,1)}$. Everywhere further we shall denote it as~$W_3$; to distinguish different dictionaries, we shall use an upper index in square brackets:~$W_3^{[j]}$, so that $f^{[j]}_{\omega} \in W_3^{[j]}$. Here $f^{[j]}_{\omega}$ is the frequency of a triplet~$\omega$. Well known Euclidean metrics
\begin{equation}\label{evklid}
\rho\left(W_3^{[1]}, W_3^{[2]}\right) = \sqrt{\sum_{\omega=\mathsf{AAA}}^{\mathsf{TTT}}\bigg(f_{\omega}^{[1]}-f_{\omega}^{[2]}\bigg)^2}
\end{equation}
has been used to determine a distance between two triplet frequency dictionaries~$W_3^{[1]}$ and~$W_3^{[2]}$, for clustering and visualization purposes.
\begin{figure}[t]
\includegraphics[width=0.49\textwidth]{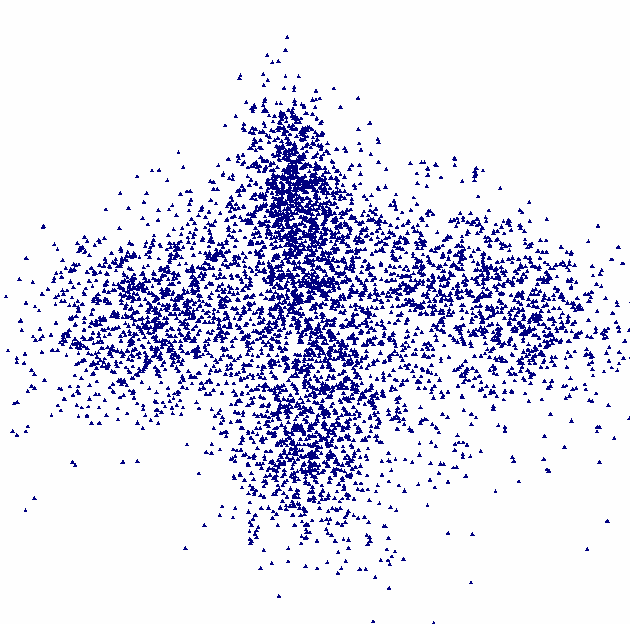}
\caption{\label{totalsimp}The total transcriptome distribution, four clusters detected.}
\end{figure}

Using \textsl{ViDaExpert} software, we considered the distribution of points corresponding to frequency dictionaries in three-dimensional projection; the choice of axes for the projection was carried out automatically, since we observed the distribution in three principal components (the first one, the second one, and the third one), mainly, not in triplets.

To prove (or disprove) visually observed clustering, we used $K$-means, provided by the same software. The choice of~$K$ was determined by the stability of clustering: we always started from $K=2$ and finished at $K^\star$ where clustering became unstable. Detail discussion of that point see in Sec.~\ref{clustering}.

Besides, we also used elastic map technique, for the purposes of visualization, mainly. Detailed description of that methodology could be found in \cite{sim2,gusev,DBLP:conf/dsaa/GorbanZ15,1742-6596-490-1-012081,10.1007/978-3-642-38679-4_50,doi:10.1142/S0129065710002383}.

\subsection{Chargaff's parity discrepancy}
Chargaff's parity rules stipulate several fundamental properties of nucleotide sequences describing a kind of symmetry in them. We used these rules to analyze the observed cluster patterns, in transcriptomes. Tot begin with, Chargaff's substitution rule stipulates that in double stranded DNA molecule nucleotide $\mathsf{A}$ always opposes to nucleotide $\mathsf{T}$, and vice versa. Same is true for the couple of nucleotides $\mathsf{C}\Leftrightarrow \mathsf{G}$.

The first Chargaff's parity rule stipulates that the number of $\mathsf{A}$'s matches the number of $\mathsf{T}$'s with a good accuracy, when counted over a single strand; obviously, similar proximal equity is observed for $\mathsf{C}$'s and $\mathsf{G}$'s. Finally, the second Chargaff's parity rule stipulates a proximal equity of frequencies of the strings comprising complementary palindrome: $f_{\omega} \approx f_{\overline{\omega}}$. Here $\omega$ and $\overline{\omega}$ are two strings counted over the same strand, so that they are read equally in opposite directions, with respect to the substitution rule, e.\,g., $\mathsf{CTGA}\Leftrightarrow \mathsf{TCAG}$; see \cite{patterns-biosys,Morton5123,Forsdyke-bioinformatics2002,MITCHELL200690,SOBOTTKA2011823,NIKOLAOU200634} for details.

Genomes differ in the figures of discrepancy of the second Chargaff's parity rule \cite{ALBRECHTBUEHLER201220}; same is true for various parts of a genome. Thus, one can compare the transcriptomes in terms of this discrepancy. To do it, let's introduce that former:
\begin{equation}\label{charg1}
\mu\left(W_3^{[1]}, W_3^{[2]}\right) = \dfrac{1}{64}\sqrt{\sum_{\omega = \mathsf{AAA}}^{\mathsf{TTT}} \Big(f^{[1]}_{\omega} - f^{[2]}_{\overline{\omega}}\Big)^2}\,,
\end{equation}
where $\omega$ and $\overline{\omega}$ are two triplets comprising complementary palindrome. Here we must take into account both couples: $f^{[1]}_{\omega} - f^{[2]}_{\overline{\omega}}$ and $f^{[2]}_{\omega} - f^{[1]}_{\overline{\omega}}$, since they exhibit different figures, in general.

Formula~\eqref{charg1} measures a deviation between two frequency dictionaries; thus, one may expect that two dictionaries $W_3^{[1]}$ and $W_3^{[2]}$ may comprise the triplets from the opposite strands, if $\mu \rightarrow 0$. An inner discrepancy measure determined within a dictionary is another important characteristics of a dictionary. To measure it, one should change the formula~\eqref{charg1} for
\begin{equation}\label{charg2}
\xi\left(W_3\right) = \dfrac{1}{32}\sqrt{\sum_{\omega \in \Omega^{\ast}} \Big(f_{\omega} - f_{\overline{\omega}}\Big)^2}\,,
\end{equation}
where $\Omega^{\ast}$ is the set of 32 couples of triplets comprising complementary palindromes. Obviously, here $|f_{\omega} - f_{\overline{\omega}}| \equiv |f_{\overline{\omega}} - f_{\omega}|$.

We shall use the figures determined by~\eqref{charg1} and~\eqref{charg2} for transcriptome analysis.

\section{Transcriptome cluster pattern}\label{clustering}
We start the analysis of transcriptomes from consideration of (real) total transcriptome (Sec.~\ref{totaltr}). Next, we consider the tissue specific transcriptomes (Sec.~\ref{spec}), and finally we present clustering patterns observed in surrogate total transcriptome (Sec.~\ref{surrog}).
\begin{figure*}
\includegraphics[width=0.49\textwidth]{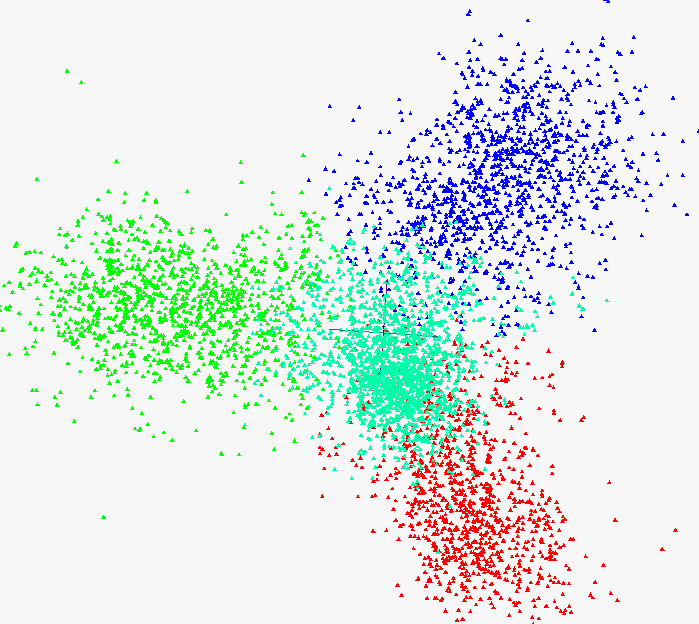}\hfill\includegraphics[width=0.49\textwidth]{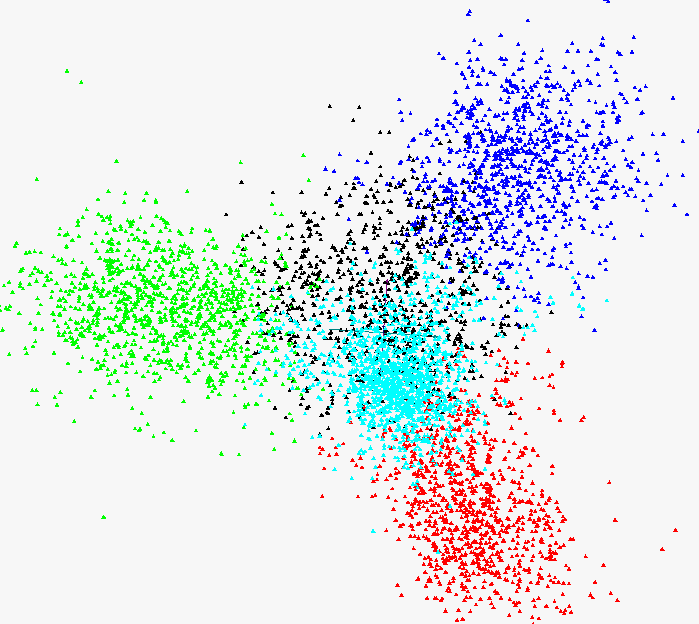}
\caption{\label{total4kl}The total transcriptome distribution, four clusters detected with $K$-means, as $K=4$ (left), and five clusters detected with $K$-means, as $K=5$ (right).}
\end{figure*}

\subsection{Total transcriptome}\label{totaltr}
The total transcriptome of \textit{L.\,sibirica} contains 610\,954 contigs; the pre-assembling filtration eliminated all the reads shorter than 300 nucleotides. To analyze the transcriptome, we selected 5\,713 longer contigs (those exceeding $5\times 10^3$ nucleotides in length); see Table~\ref{dl}. To analyze the transcriptome, we have converted each selected contig into triplet frequency dictionary~$W_3^{[j]}$, and studied the distribution of the relevant points in 64-dimensional metric space.

Fig.~\ref{totalsimp} shows the distribution of the contigs. Apparently, there are four clusters in the distribution. It should be stressed that the clusters shown in Fig.~\ref{totalsimp} still require an objective proof. In other words, the visualization yields four clusters, but whether they could be identified by other methods or techniques independent on a researcher's sight, still must be proven (or disproved).

Speaking on $K$-means clustering, one should keep in mind the problem of stability of the obtained clustering \cite{dyn1}. It should be stressed, that the clustering is very unstable, for $K=2$ and $K=3$. The problem arises from the random starting distribution of points, in $K$-means clustering. A pattern observed for some specific $K$ is called \textbf{\textbf{stable}}, if it occurs in a given portion of final distributions obtained in a series of runs of the $K$-means clustering. Of course, the portion level~$\delta$ is a matter of choice of a researcher; usually, $\delta = 0.8$ is supposed to be the high stability figure.

We checked the stability of clustering in a series of a hundred of runs of $K$-means. A distribution is claimed to be stable, if the configuration occurs in 80 runs, in the series of a hundred. We have developed the clustering for $2 \leq K \leq 5$. The clustering for $K=2$ and $K=3$ are highly unstable. On the contrary, the clustering for $K=4$ yields very high stability. Surprisingly, clustering for $K=5$ is also very stable, and identified four previously visible cluster, plus an intermediate space between them, as a separate cluster.

The evidence for the clear and unambiguous cluster identification is shown in Fig.~\ref{total4kl}. Here the left figure shows the distribution of contigs converted in the points in 64-dimensional metric space obtained by $K$-means technique, with $K=4$. The right figure shows the same distribution, while $K=5$.

Green, red, dark blue and light blue colors mark up the same clusters (in general), and black shows the fifth cluster appeared due to $K$-means clustering with $K=5$. Obviously, the points comprising the fifth cluster had come in it from four others. Nonetheless, both clusterings (for $K=4$ and $K=5$) are highly stable: ninety runs of $K$-means converts to the same configuration of the points, in the series of a hundred of runs. The most remarkable fact here is that the fifth cluster colored in black occupies an intermediate space, between the four ones clearly identified for $K=4$.

\subsubsection{Chargaff's figures for total transcriptome}
We hypothesize that four clusters observed for total transcriptome correspond to the different tissues involved in the total transcriptome implementation. Yet, we have neither proof, nor disproof for that hypothesis; the bruit force method to prove it is to map reads back on contigs, and identify the tissue attribution of a contig. Another approach is significantly less costly, and may bring an approximate answer. To get it, one should find the figures~\eqref{charg1} and~\eqref{charg2} for the centers of the classes identified through $K$-means implementation to the set of contigs.
\begin{figure*}
\includegraphics[width=0.49\textwidth]{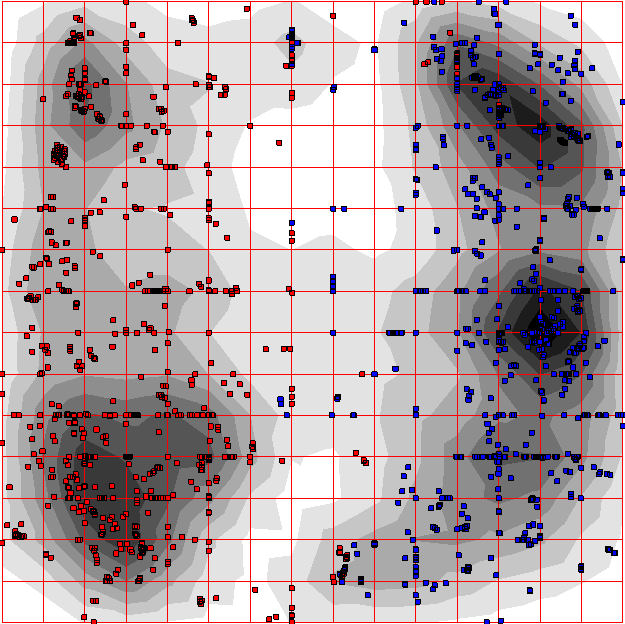}\hfill\includegraphics[width=0.49\textwidth]{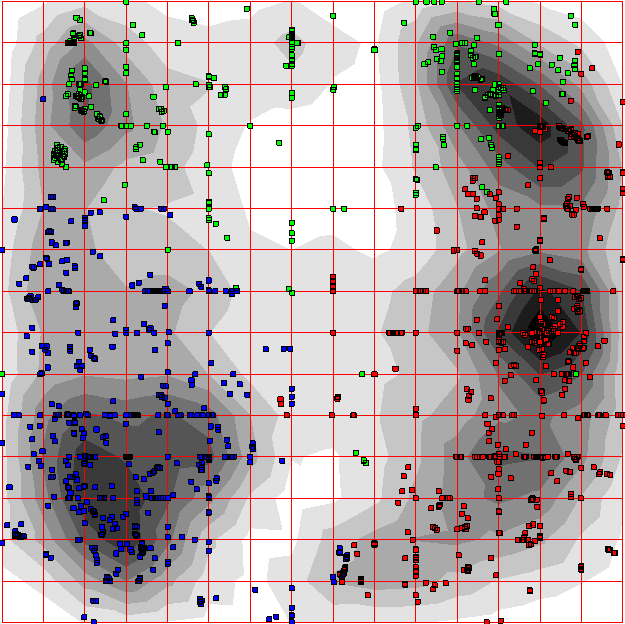}\\[6pt]
\includegraphics[width=0.49\textwidth]{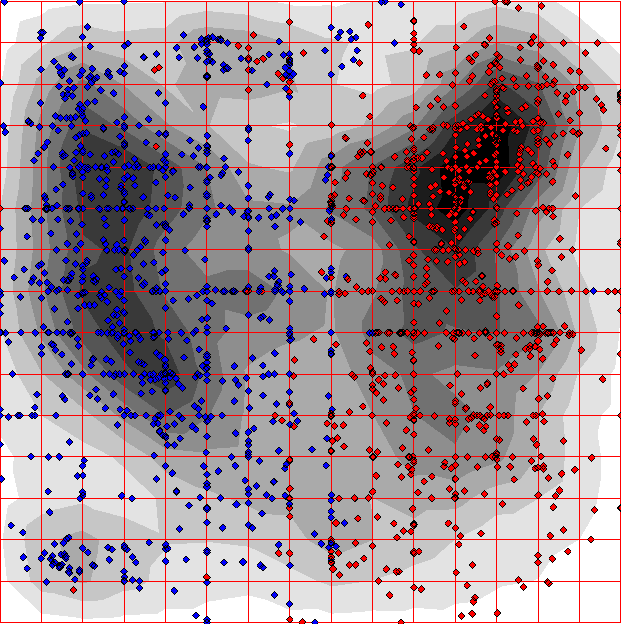}\hfill\includegraphics[width=0.49\textwidth]{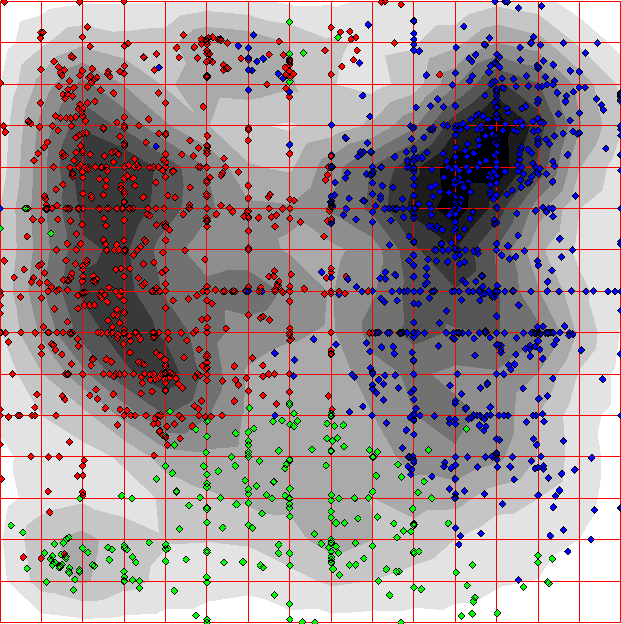}\\
\caption{\label{specific}Transcriptomes of cambium (upper) and needles (lower). Left picture shows two cluster pattern, and the right one shows three cluster pattern.}
\end{figure*}

Similar approach has been used when studied the larch transcriptome \cite{sfu-bio-trans,sfu-bio-chloro}, and the figures effectively identified bacterial contamination, and strand allocation of the contigs comprising various classes. Here we explore similar approach, and determined the discrepancy~$\xi$ for those four clusters shown in Figs.~\ref{totalsimp} and~\ref{total4kl}: $\xi_1 = 0.001921$, $\xi_2 = 0.001368$, $\xi_3 = 0.001626$ and $\xi_4 = 0.001615$. Here $\xi_j$ is the discrepancy value observed for $j$-th class (see Eq.~\eqref{charg2}).

Similar figures $\mu_{\{i,j\}}$ for interclass discrepancy (see Eq.~\eqref{charg1}) are: $\mu_{\{1,2\}} = 0.000618$, $\mu_{\{1,3\}} = 0.000933$, $\mu_{\{1,4\}} = 0.000668$, $\mu_{\{2,3\}}=0.000731$, $\mu_{\{2,4\}} = 0.000522$ and $\mu_{\{3,4\}} = 0.000663$. It should be stressed that both~\eqref{charg1} and~\eqref{charg2} characteristics were determined over the dynamic kerns of the clusters; these are the arithmetic mean of the triplet frequencies determined over the set of contigs enlisted into a cluster.
\begin{figure*}
\includegraphics[width=0.49\textwidth]{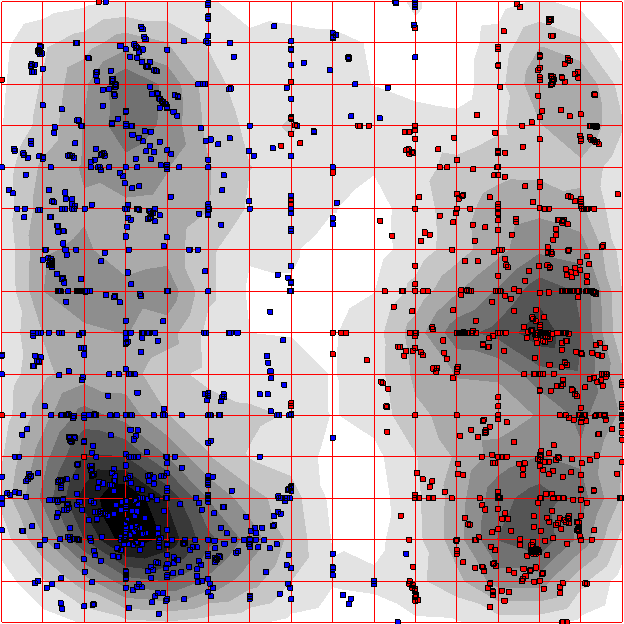}\hfill\includegraphics[width=0.49\textwidth]{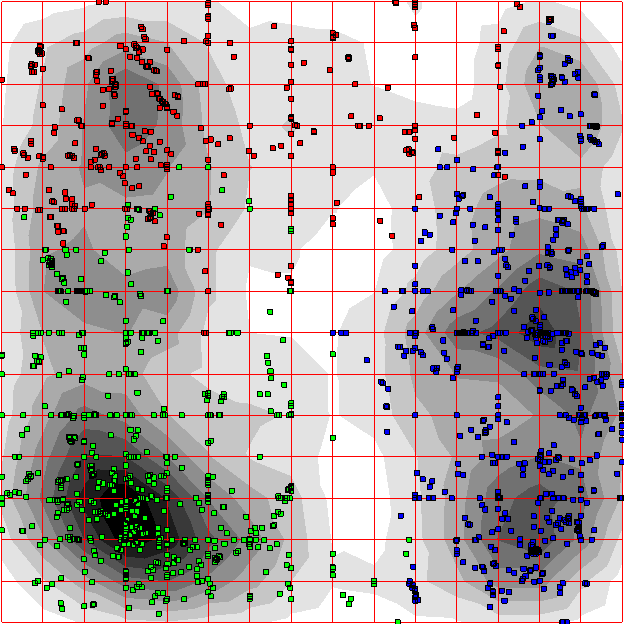}\\[6pt]
\includegraphics[width=0.49\textwidth]{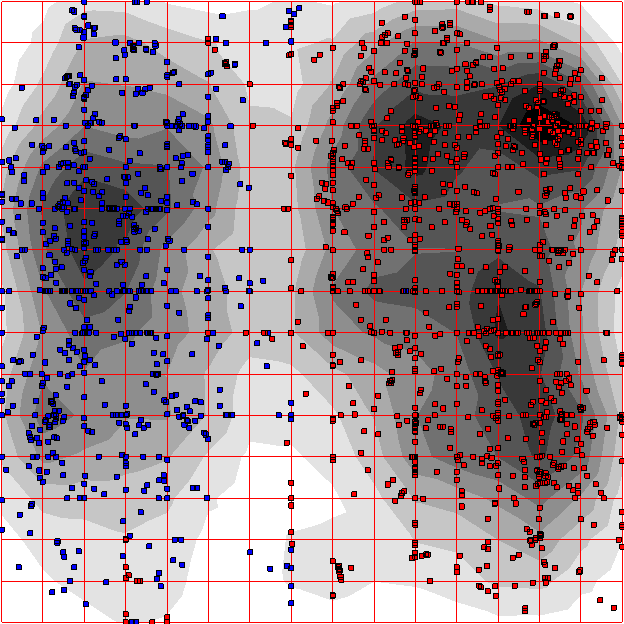}\hfill\includegraphics[width=0.49\textwidth]{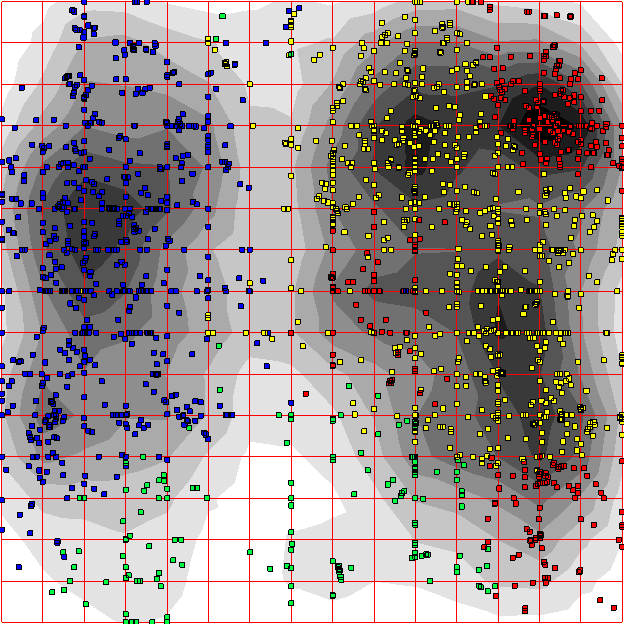}\\
\caption{\label{specific1}Transcriptomes of shoot (upper) and seedlings (lower). Left picture shows two cluster pattern, and the right one shows three cluster pattern (for shoot) and four cluster pattern (for seedlings).}
\end{figure*}

\subsection{Tissue specific transcriptomes}\label{spec}
Besides the total transcriptome, four tissue specific transcriptomes have been assembled: needles, shoot, seedlings and cambium. The figures characterizing these transcriptomes are shown in Table~\ref{dl}. It should be stressed that these transcriptomes differ in abundances: cambium transcriptome is the richest one, while the needles transcriptome is the lowest. To avoid an abundance bias impact, we balanced the parts of transcriptomes taken into consideration so that $M_d$ figures are quite close. To do it, we have to include into the subsets the contigs of different length; since the lower boundary of the cut-off length is 1\,000 nucleotides, one may expect no finite sampling effects in the contigs (converted into frequency dictionaries) distribution.

We shall present the clustering results for tissue specific transcriptome with the help of elastic map. $16\times 16$ cell soft elastic map has been implemented, for all four tissue specific transcriptomes. We used the standard software settings, for map adjusting. The distributions for all four transcriptomes are shown in so called inner coordinates (see \cite{sim2,gusev,DBLP:conf/dsaa/GorbanZ15,1742-6596-490-1-012081,10.1007/978-3-642-38679-4_50,doi:10.1142/S0129065710002383} for details). Grey scaled contours show the average local density of the points (these are triplet frequency dictionaries); ten grade scale was used. Obviously, the darker is background, the higher is local density, and vice versa.
\begin{figure}[t]
\includegraphics[width=0.49\textwidth]{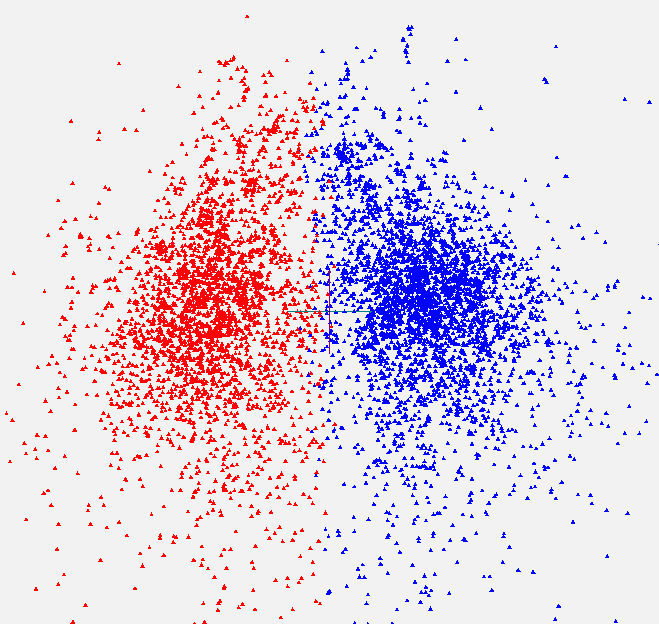}
\caption{\label{surrogat1}Total surrogate transcriptome distribution, two clusters detected.}
\end{figure}

The tissue specific transcriptome distributions are shown in Figs.~\ref{specific} and~\ref{specific1}. All eight pictures show the stable clustering. In particular, the $K$-means clustering into two and three classes was stable for needles, shoot, and cambium transcriptomes. Surprisingly, the seedlings transcriptome yields very low stability when clustered by $K$-means for three clusters, and reasonable stability when clustered for four clusters. This fact is not extremely seldom, while it says on the increased complexity of that transcriptome, in comparison to three others.

\subsection{Surrogate total transcriptome}\label{surrog}
Finally, we generated a surrogate total transcriptome to check our hypothesis. It should be stressed that the study of a surrogate transcriptome may not be a proof (or disproof), for the hypothesis. Meanwhile, it is an indirect evidence, pro or contra.

The surrogate total transcriptome was developed through an immediate merging of all four tissue specific transcriptomes, in a single one. Developing the surrogate transcriptome, we kept the information on the tissue origin of each contig. Thus, we could trace the distribution of the contigs belonging to different tissues over the pattern observed for surrogate total transcriptome. Four tissue specific transcriptomes yield 7003 contigs into a surrogate one; this figure exceeds the abundance of the real total transcriptome contigs subset converted into triplet frequency dictionaries, while the excess is not significant.

To begin with, the surrogate transcriptome differs significantly from the real total one. There are no four-cluster structure in the surrogate transcriptome; instead, the two-cluster one is observed, see Fig.~\ref{surrogat1}. No regularity in tissue distribution over the clusters is observed: indeed, the first cluster contains 674 points belonging to \textsl{cambium}, 942 points belonging to \textsl{seedlings}, 980 points belonging to \textsl{needles}, and 1228 points belonging to \textsl{shooters}. Reciprocally, the second cluster contains 781, 812, 871, 715 points from the same tissues, respectively. All the tissues but shooters are distributed between the clusters almost equally; the shooters contigs exhibit a kind of preference in the cluster occurrence: the first cluster contains almost twice points more, than the second one. This fact might support indirectly the original hypothesis on the tissue speciality of the clusters observed in the original total transcriptome.

\section{Discussion}
Here we present some very preliminary results on the analysis of statistical properties of Siberian larch (total) transcriptome. Total transcriptome analysis is an essential step in deciphering any genome (see e.\,g. \cite{Wang2016}); the basic idea standing behind it is to enhance the quality of assembling through a kind of improvement of coverage depth, in sequencing. Indeed, a sequencing may yield low (or insufficient) coverage level, thus making an assembling rather arguable, or unreliable. Since total transcriptome is expected to be enriched with the subensembles of reads sequenced from the same genes expressing in different tissues, then low coverage in one tissue sample may be compensated by high (or at least reasonable) coverage level obtained in other tissue samples.

At the first glance, this idea seems to be quite fruitful. The point is that such hypothesis is based on a na\"{\i}ve expectation of the total equivalence of the reads gained from different tissue samples. This equivalence is expected to be come from the identity of genes family expressed in various tissues; at least, a part of genes do express in various tissues, simultaneously. The point is the greater difference in $k$-mers ending the reads: an effective enrichment (and, hence, improvement of assembling) of coverage level will be effective, iff the set of possible $k$-mers fixed at the ends of reads to carry out a graph implementation, does not grow up, or at least, grows very slowly.

This expectation does not hold true, in general \cite{Wang2016}. A lack of unambiguity in $k$-mers set detected in total transcriptome seems to be a common place \cite{Shi2016,transcranalysisLily2016}. The situation is easier, for animals \cite{CHEN2015723}, while yet it does not solve the problem completely. Nonetheless, a study of total transcriptome may bring some important and valuable knowledge on some details of the specific transcriptomes, as well as on the actively expressing genetic system.

The brute force method here implies a hard and labour-consuming alignment; that latter is not so labourious, when one is able to refer to a reference genome. This situation becomes very hard, for \textsl{de novo} genome (or transcriptome) assembling. Various statistical methods and techniques takes an advantage, in this case. The simplest approach is to find out whether the genetic entities under consideration possess an order. Obviously, the answer strongly depends on the idea of an order. Here clustering (with some proper technique) may answer in the simplest way this question.

There exists a tremendous variety of clustering methods, and $K$-means is number one among them. Paper \cite{kim2017} presents the results of the order identification, in a transcriptome. This paper presents the clustering approach to identify the reads with close $k$-mers, in order to carry our further assembling. Since the authors consider the $k$-mers as is, they have to deal with memory constraints problem, as well as some computational problems. Also, the paper presents the biologically sounding results of $K$-mean application.

The practice of clustering analysis for genetic sequences investigation is quite wide; in particular, transcriptome analysis of various species has been carried out and presented in \cite{Bedre2016,Devani2017,denovoForsythia,denovoSorbus}; some special issues of statistical methods implementation are considered in \cite{stattrainngcoding2013}.

A study of total transcriptome is an essential step, in differential expression investigations. Indeed, a study of transcriptomes of some specific tissues, or under some peculiar conditions might be strongly supported by the investigation of the total transcriptome, which serves here as a kind of a reference entity. For instance, an environmental impact on transcriptome composition and statistical properties is studied in \cite{Zhang2017}; sexual variability of a transcriptome is studied and discussed in \cite{genes8120393,Qiao2016}.

Our data present in this paper are in general concord to the observations and findings presented in the discussed papers. Further studies will be targeted on:
\begin{list}{\arabic{nnn})}{\usecounter{nnn}\leftmargin=6mm \labelwidth=5mm \topsep=0mm \labelsep=2mm \itemsep=1pt \parsep=0mm \itemindent=10pt}
\item the most detailed study of the four-cluster pattern observed due to $K$-means, in our total transcriptome;
\item detailed studies of inner structuredness of tissue specific transcriptomes;
\item comparative studies of the structure entities identified in tissue specific transcriptomes.
\end{list}

We suppose to carry out the mapping of the reads over the total transcriptome contigs, to figure out the tissue difference impact into the total transcriptome, thus proving or disproving the hypothesis on the origin of four cluster patter of that former. All these results will be present in the nearest papers.

\bibliography{guseva}
\end{document}